\documentclass{sig-alternate}
\usepackage{algorithmic,algorithm}
\begin{document}
%
\title{SART: Speeding up Query Processing in Sensor Networks with an Autonomous Range Tree Structure}

\numberofauthors{4} 
\author{
\alignauthor
Spyros Sioutas \\
  \affaddr{Department of Informatics, Ionian University}\\
  \affaddr{49100 Corfu, Greece}\\
  \email{sioutas@ionio.gr}
  \alignauthor
Alexandros Panaretos \\
  \affaddr{Department of Informatics, Ionian University}\\
  \affaddr{49100 Corfu, Greece}\\
  \email{alex@ionio.gr}
  \alignauthor
Ioannis Karydis \\
  \affaddr{Department of Informatics, Ionian University}\\
  \affaddr{49100 Corfu, Greece}\\
  \email{karydis@ionio.gr}
  \and
\alignauthor
Dimitrios Tsoumakos \\
  \affaddr{Department of Informatics, Ionian University}\\
  \affaddr{49100 Corfu, Greece}\\
  \email{dtsouma@ionio.gr}
\alignauthor
Giannis Tzimas \\
  \affaddr{Dept. of Applied Informatics in Management and Economy, Techn. Educ. Institute}\\
  \affaddr{30200 Messolonghi, Greece}\\
  \email{tzimas@teimes.gr}
\alignauthor
Dimitrios Tsolis \\
  \affaddr{Cult. Herit. Management and New Technologies Dept., University of Western Greece}\\
  \affaddr{30100 Agrinio, Greece}\\
  \email{dtsolis@upatras.gr}
}

\maketitle
\begin{abstract}
We consider the problem of constructing efficient P2P overlays for sensornets providing "Energy-Level Application and Services". In this context, assuming that a sensor is responsible for executing some program task but unfortunately it's energy-level is lower than a pre-defined threshold.  Then, this sensor should be able to introduce a query to the whole system in order to discover efficiently another sensor with the desired energy level, in which the task overhead must be eventually forwarded. In this way, the "Life-Expectancy" of the whole network could be increased. Sensor nodes are mapped to peers based on their energy level. As the energy levels change, the sensor nodes would have to move from one peer to another and this operation is very crucial for the efficient scalability of the proposed system. Similarly, as the energy level of a sensor node becomes extremely low, that node may want to forward it's task to another node with the desired energy level. The method presented in \cite{SOPXM09} presents a novel P2P overlay for Energy Level discovery in a sensornet. However, this solution is not dynamic, since requires periodical restructuring. In particular, it is not able to support neither join of sensor\_nodes with energy level out of the ranges supported by the existing p2p overlay nor leave of \emph{empty} overlay\_peers to which no sensor\_nodes are currently associated. On this purpose and based on the efficient P2P method presented in \cite{SPSTMT10}, we design a dynamic P2P overlay for Energy Level discovery in a sensornet, the so-called SART (Sensors' Autonomous Range Tree) \footnote{Preliminary version of this work was presented in ACM SAC'12, pp.847-852, March 25-29, 2012, Riva del Garda, Italy}. The adaptation of the P2P index presented in \cite{SPSTMT10} guarantees the best-known dynamic query performance of the above operation. We experimentally verify this performance, via the D-P2P-Sim simulator \footnote{\tt D-P2P-Sim is publicly available at http://code.google.com/p/d-p2p-sim/}.  
\end{abstract}




\section{Introduction}

In the last years sensornet research primarily focused on data collection, finding applications in ecology (e.g., environmental and habitat monitoring \cite{MPSCA02}), in precision agriculture (e.g., monitoring of temperature and humidity), in civil engineering (e.g., monitoring stress levels of buildings under earthquake simulations), in military and surveillance (e.g., tracking of an intruder \cite{DAG03}), in aerospace industry (e.g., fairing of cargo in a rocket), etc.\\

Traditionally, sensors are used as data gathering instruments, which continuously feed a central base station database. The queries are executed in this centralized  base  station  database  which  continuously collates the data. However, given the current trends (increase in numbers of sensors, together collecting gigabits of data, increase in processing power at sensors) it is not anymore feasible to use a centralized solution for querying the sensor networks. Therefore, there is a need for establishing an efficient access structure on sensor networks in order to contact only the relevant nodes for the execution of a query and hence achieve minimal energy  consumption, minimal response time, and an accurate response. We achieve these goals with our peer-to-peer query processing model on top of a distributed index structure on wireless sensor networks.\\

In sensor networks any node should be able to introduce a query to the system. For example, in the context of a fire evacuation scenario a firefighter should be able to query a nearby sensor node for the closest exit where safe paths exist.  Therefore, a peer-to-peer query processing model is required. A first P2P program for spatial query execution presented in \cite{DF07}. \\

According to \cite{AL07}, the benefits of the P2P overlays in sensornets are the following:
Efficient Data Lookup, Guaranties on Lookup Times, Location Independence, Overlay Applications and Services, Elimination of proxies/sinks with undesirable central authority, Limited Broadcast. P2P design, for Internet-like environments, has been a very active research area and there are many P2P Internet protocols and systems available like CAN \cite{BYL08}, Pastry \cite{BYL08}, and Chord \cite{BYL08}. The main arguments against P2P designs in sensornets were the following: Logical Topology=Physical Topology, Route Maintenance Overhead, Sensor Nodes are Not Named, DHTs are Computationally Intensive.
By overcoming the arguments above (for details see \cite{AL07}, \cite{AU04} and \cite{CCNOR06}), in \cite{AU04} and \cite{CCNOR06} the first DHT (Distributed Hash Table) based protocols for sensornets were presented, the CSN and VRR respectively. In \cite{AL07} the Tiered Chord (TChord) protocol was proposed, which is similar to, and inspired by, CSN. TChord is a  simplified  mapping  of Chord onto  sensornets. Unlike CSN the design of TChord is more generic (to support a variety of applications and services on top instead of just serving incoming data queries). Gerla et  al.  argue for the  applicability  and  transfer of wired P2P models and techniques to MANETs \cite{GLR05}. \\
Most existing decentralized discovery solutions in practice are either DHT based, like Chord or hierarchical clustering based, like BATON \cite{BYL08}, NBDT \cite{S08}, ART \cite{SPSTMT10} or Skip-Graphs \cite{BYL08}.
The majority of existing P2P overlays for sensornets were designed in a DHT fashion and the best current solution is the TChord.
On the contrary, ELDT \cite{SOPXM09} is the only existing P2P protocol for sensornets, which combines the benefits of both DHT and hierarchical \cite{S08} clustering fashions. In this solution, sensor\_nodes are mapped to peers based on their energy level. As the energy levels change, the sensor nodes would have to move from one peer to another and this oparation is very crucial for the efficient scalability of the proposed system. Similarly, as the energy level of a sensor node becomes extremely low, that node may want to forward it's task to another node with the desired energy level. However, the ELDT solution is not dynamic, since requires periodical restructuring. In particular, it is not able to support neither join of sensor\_nodes with energy level out of the ranges supported by the existing p2p overlay nor leave of \emph{empty} overlay\_peers to which no sensor\_nodes are currently associated. On this purpose and based on the efficient P2P method presented in \cite{SPSTMT10}, we design a dynamic P2P overlay for Energy Level discovery in a sensornet, the so-called SART (Sensors' Autonomous Range Tree). The adaptation of the P2P index presented in \cite{SPSTMT10} guarantees the best-known dynamic query performance of the above operation.

The main functionalities of SART attempt to increase the "Life-Expectancy" of the whole sensor network in dynamic way, providing support for processing: (a) exact match queries of the form "given a sensor node with low energy-level $k'$, locate a sensor node with high energy-level $k$, where $k>>k'$" (the task will be forwarded to the detected sensor node) (b) range queries of the form "given an energy-level range $\left[k, k'\right]$, locate the sensor node/nodes the energy-levels of which belong to this range" (the task will be forwarded to one of the detected sensor nodes) (c) update queries of the form "find the new overlay-peer to which the sensor node must be moved (or associated) according to it's current energy level" (the energy level of each sensor node is a decreasing function of time and utilization) (d) join queries of the form "join a new overlay-peer to which the new (inserted) sensor node is associated" and (e) leave queries of the form "leave (delete) the overlay-peer to which no sensor nodes are currently associated". The SART overlay adapts the novel idea of ART P2P infrastructure presented in \cite{SPSTMT10} providing functionalities in optimal time. For comparison purposes, an elementary operation's evaluation is presented in table 1 between ART, NBDT, Skip-Graphs \cite{BYL08}, Chord \cite{BYL08} and its newest variation (F-Chord(á) \cite{BYL08}), BATON and its newest variation (BATON* \cite{BYL08}).
The rest of this paper is structured as follows. Section 2 and 3 describe the SART system while section 4 presents an extended experimental verification via an appropriate simulator we have designed for this purpose. Section 5 concludes.
 
\begin{table*}[htbp]
	\begin{center}
		\begin{tabular}{| c | c | c |c | c |}
		 \hline
		  P2P Architectures & Lookup/update key & Data Overhead-Routing information & Join/Depart Node \\ \hline
			Chord & $O(logN)$  & $O(logN)$ nodes & $O(logN)$ w.h.p. \\ \hline
			H-F-Chord(a)  & $O(log N/ log log N)$ & $O(logN)$ nodes & $O(logN)$ \\ \hline
			LPRS-Chord & $O(logN)$ & $O(logN)$ nodes & $O(logN)$ \\ \hline
			Skip Graphs & $O(logN)$ & $O(1)$ & $O(logN)$ amortized \\ \hline
			BATON & $O(logN)$ & Two (2) nodes & $O(logN)$ w.h.p. \\ \hline
			BATON* & $O(log_{m}N)$ & $m$ nodes & $O(mlog_{m}N)$ \\ \hline
			NBDT & $O(loglogN)$ & $O(loglogN)$ or $2^{2^{i-1}}$ for nodes at level $i$ of left spine 
			& periodical restructuring \\ \hline
			ART & $O(\log_{b}^2 \log N)$ & $O(N^{1/4}/ \log^c N)$ nodes & $O(\log \log N)$ expected w.h.p. \\ \hline
		\end{tabular}
	\end{center}
	\caption{Performance Comparison between ART, NBDT, Chord, BATON and Skip Graphs}
	\label{tab:perf_comp}
\end{table*}

\section{The SART protocol}

SART, is a simplified mapping of ART \cite{SPSTMT10} onto sensornets. Like ART, at the heart of SART, lookup and join/leave respectively are the two main operations. Given a set of sensor nodes, we hash the unique address of each sensor node to obtain node identifiers. Meta-data keys, generated from the data stored on the nodes, are hashed to obtain key identifiers.
 
\begin{figure}[htbp]
	\begin{center}
		\includegraphics[scale=0.40]{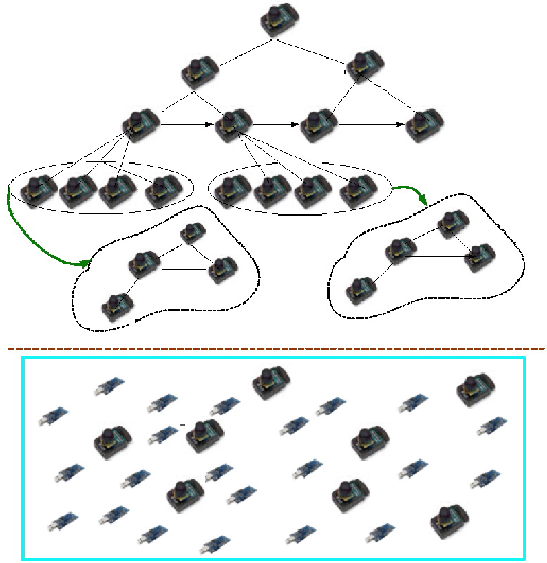}
	\end{center}

	\caption{The SART protocol}
	\label{fig:master_slave_sart}
\end{figure}

\begin{figure}[htbp]
	\begin{center}
		\includegraphics[scale=0.50]{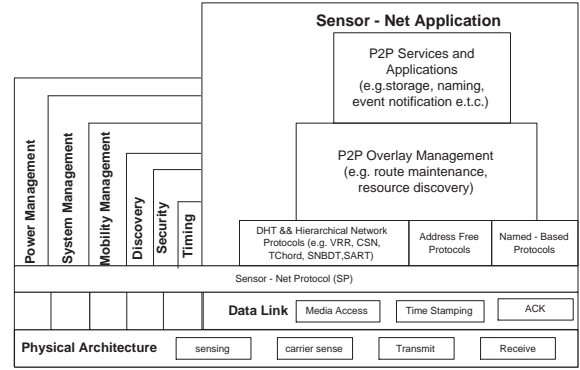}
	\end{center}

	\caption{P2P Overlay in SP Architecture}
	\label{fig:SP_Architecture}
\end{figure}

The SART protocol (see figure \ref{fig:master_slave_sart}) is an hierarchical arrangement of some sensor nodes (master nodes).
The master node of level $i$ maintains information (in its local finger table) about all its \emph{slave nodes} and  $2^{2^{i-1}}$ other master nodes (you can find more details about master and slave nodes in \cite{SOPXM09}). All queries are resolved in a distributed manner with a bound of $O(\log_{b}^2 \log N)$ messages. When a master node receives a query it first checks its own keys to resolve the query, if the lookup is not successful the master node then checks its local finger table. The finger table contains information about $2^{2^{i-1}}$ other master nodes and if the key can be located according to the information stored in the finger table, the query is directly forwarded to the master node storing the data. If the lookup on the local finger table also fails then the master node routes the query to the master node closest to the target according to the finger table.
We handle the master\_node joins/leaves and fails according to join/leave and fail operations respectively presented in \cite{SPSTMT10}. Thus, all the above operations are bounded by $O(\log \log N)$ expected w.h.p. number of messages. 
Slave nodes do not store information about their neighbors. If a slave node directly receives a query, it checks its own data and if the lookup fails it simply forwards the query to its master node. For simplicity, in the SART proposal we opt for not connecting the slave nodes in a ART arrangement and lookups are not implemented in slave nodes. The master nodes could be thought as "virtual sinks" with an ART overlay between these virtual sinks.
Unlike IP in the Internet, the sensornet protocol SP is not at the network layer but instead sits between the network and data-link layer (because data-processing potentially occurs at each hop, not just at end points).
Figure \ref{fig:SP_Architecture} shows how P2P overlays can be implemented on top of SP. The P2P overlay (shown as P2P Overlay Management) could be built on top of any generic network protocol. An underlying DHT or Hierarchical Clustering routing protocol (e.g., VRR, CSN, TChord or SNBDT or SART) is recommended as it simplifies the job of overlay management. In particular, it is more efficient to build routing directly on top of the link layer instead of implementing it as an overlay on top of a routing protocol \cite{CCNOR06}. P2P Services and Applications (e.g.  event notification, resource allocation, and file systems) can then be built on top of the P2P overlay and sensornet applications could either use these services or communicate with the P2P overlay themselves.

\section{The SART P2P overlay}

Let $G$ a network graph of $n$ sensor nodes and SART the respective overlay of $N$ peers. With each overlay peer $p$ 
($1\leq p\leq N$) we associate a set of pairs $S_p=\left\{(g,L[g])\right\}$, where $g$ is a sensor node 
($1\leq g \leq n$) and $L[g]$ its current energy level. The criterion of associating the sensor node $g$ to peer $p$ depends on it's current energy level. Obviously, it holds that $N<<n$. Let's explain more the way we structure our whole system.

One of the basic components of the final SART structure is the LRT (\textbf{L}evel \textbf{R}ange \textbf{T}ree) \cite{SPSTMT10} structure. LRT will be called upon to organize collections of peers at each level of SART.

\subsection{The LRT structure \cite{SPSTMT10}: An overview} 
\begin{figure}[htbp]
	\centering
		\includegraphics[width=0.44\textwidth]{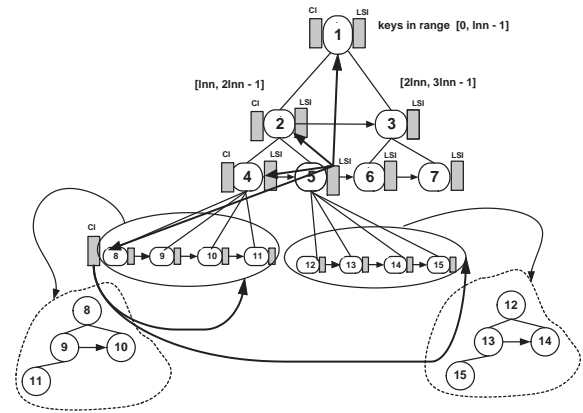}
	\caption{The LRT structure}
	\label{fig:LRT}
\end{figure}

LRT is built by grouping nodes having the same ancestor and organizing them in a tree structure recursively. The innermost level of nesting (recursion) will be characterized by having a tree in which no more than $b$ nodes share the same direct ancestor, where $b$ is a double-exponentially power of two (e.g. 2,4,16,...). Thus, multiple independent trees are imposed on the collection of nodes. Figure~\ref{fig:LRT} illustrates a simple example, where $b=2$. 

The degree of the overlay peers at level $i>0$ is $d(i)=t(i)$, where $t(i)$ indicates the number of peers at level $i$. It holds that $d$(0)=2 and $t$(0)=1. 
Let $n$ be $w$-bit keys. Each peer with label $i$ (where $1 \leq i \leq N$) stores ordered keys that belong in the range [$(i-1) \ln n , i \ln n$--1], where $N=n/lnn$ is the number of peers. Each peer is also equiped with a table named {\it Left Spine Index} (LSI), which stores pointers to the peers of the left-most spine (see pointers starting from peer 5). Furthermore, each peer of the left-most spine is equipped with a table named {\it Collection Index} (CI), which stores pointers to the collections of peers presented at the same level (see pointers directed to collections of last level). Peers having the same father belong to the same collection. 

\emph{\textbf{Lookup Algorithm:}} Assume we are located at peer $s$ and seek a key $k$. First, the algorithm finds the range where $k$ belongs. If $k \in [(j-1)$ $\ln n, j \ln n-1]$, it has to search for peer $j$. The first step of algorithm is to find the LRT level where the desired peer $j$ is located. For this purpose, it exploits a nice arithmetic property of LRT. This property says that for each peer $x$ located at the left-most spine of level $i$, the following formula holds:
\begin{equation} \label{eq:father}
label(x)=label(father(x))+2^{2^{i-2}}
\end{equation}
For each level $i$ (where $0 \leq i \leq \log \log N$), it computes the value $x$ of its left most peer by applying Equation~(\ref{eq:father}). Then, it compares the value $j$ with the computed value $x$. If $j \geq x$, it continues by applying Equation (1), otherwise it stops the loop process with current value $i$. The latter means that node $j$ is located at the $i$-th level. Then, it follows the $i$-th pointer of the LSI table located at peer $s$.
Let $x$ the destination peer, that is the leftmost peer of level $i$. Now, the algorithm must compute the collection in which the peer $j$ belongs to. Since the number of collections at level $i$ equals the number of nodes located at level $(i-1)$, it divides the distance between $j$ and $x$ by the factor $t(i-1)$ and let $m$ the result of this division. Then, it follows the $(m+1)$-th pointer of the CI table. Since the collection indicated by the CI[$m$+1] pointer is organized in the same way at the next nesting level, it continues this process recursively.
 
\emph{\textbf{Analysis:}} Since $t(i)=t(i-1)d(i-1)$, it gets $d\left(i\right)=t\left(i\right)=2^{2^{i-1}}$ for $i\geq 1$. Thus, the height and the maximum number of possible nestings is $O(\log \log N)$ and $O(\log_{b} \log N)$ respectively. Thus, each key is stored in $O(\log_{b} \log N)$ levels at most and the whole searching process requires $O(\log_{b} \log N)$ hops. Moreover, the maximum size of the $CI$ and $RSI$ tables is $O(\sqrt{N})$ and
$O(\log \log N)$ in worst-case respectively.

Each overlay peer stores tuples $(g, L[g])$, where $L[g]$ is a $k-bit$ key belonging in universe $K=[0,2^{k}-1]$, which represents the current energy-level of the sensor node $g$. We associate to $i^{th}$ peer the set $S_i=\left\{(g,L[g])\right\}$, where $L_g\in [(i-1)lnK, ilnK-1]$. Obviously, the number of peers is $N=K/lnK$ and the load of each peer becomes $\Theta (polylogN)$ in expected case with high probability (for more details see[1]). Each energy-level key is stored at most in $O(loglogN)$ levels. We also equip each peer with the table $LSI$ (Left Spine Index). This table stores pointers to the peers of the left-most spine (for example in figure $3$ the peers $1$, $2$, $4$ and $8$ are pointed by the LSI table of peer $5$) and as a consequence its maximum length is $O(loglogN)$.\\

Furthermore, each peer of the left-most spine is equipped with the table $CI$ (Collection Index). $CI$ stores pointers to the collections of peers presented at the same level (see in figure 3 the CI table of peer $8$). Peers having same father belong to the same collection. For example in the figure 2, peers $8$,$9$,$10$ and $11$ constitute a collection of peers. It's obvious that the maximum length of CI table is $O(\sqrt{N})$.

\subsection{The ART \cite{SPSTMT10} structure: An Overview} 

The backbone of ART is exactly the same with LRT. During the initialization step the algorithm chooses as cluster\_peer representatives the 1st peer, the $(\ln n)$-th peer, the $(2 \ln n)$-th peer and so on.\\
This means that each cluster\_peer with label $i'$ (where $1 \leq i'\leq N'$) stores ordered peers with energy-level keys belonging in the range $[(i'-1) \ln^2 n, \ldots, i' \ln^2 n-1]$, where $N'=n/ \ln^2 n$ is the number of cluster\_peers. 

ART stores cluster\_peers only, each of which is structured as an independent decentralized architecture. Moreover, instead of the {\bf L}eft-most {\bf S}pine {\bf I}ndex (LSI), which reduces the robustness of the whole system, ART introduces the {\bf R}andom {\bf S}pine {\bf I}ndex (RSI) routing table, which stores pointers to randomly chosen (and not specific) cluster\_peers (see pointers starting from peer 3). In addition, instead of using fat $CI$ tables, the appropriate collection of cluster\_peers can be accessed by using a 2-level LRT structure. 

\emph{\textbf{Load Balancing:}} The join/leave of peers inside a cluster\_peer were modeled as the combinatorial game of bins and balls presented in \cite{KMSTTZ03}. In this way, for a $\mu(\cdot)$ random sequence of join/leave peer operations, the load of each cluster\_peer never exceeds $\Theta({\rm polylog}~N')$ size and never becomes zero in expected w.h.p. case.

\emph{\textbf{Routing Overhead:}} The 2-level LRT is an LRT structure over $\log^{2c}Z$ buckets each of which organizes $\frac{Z}{\log^{2c}Z}$ collections in a LRT manner, where $Z$ is the number of collections at current level and $c$ is a big positive constant. As a consequence, the routing information overhead becomes $O(N^{1/4}/ \log^c N)$ in the worst case (even for an extremely large number of peers, let say N=1.000.000.000, the routing data overhead becomes $6$ for $c=1$).

\begin{figure*}
\begin{center}
\includegraphics[width=0.80\textwidth]{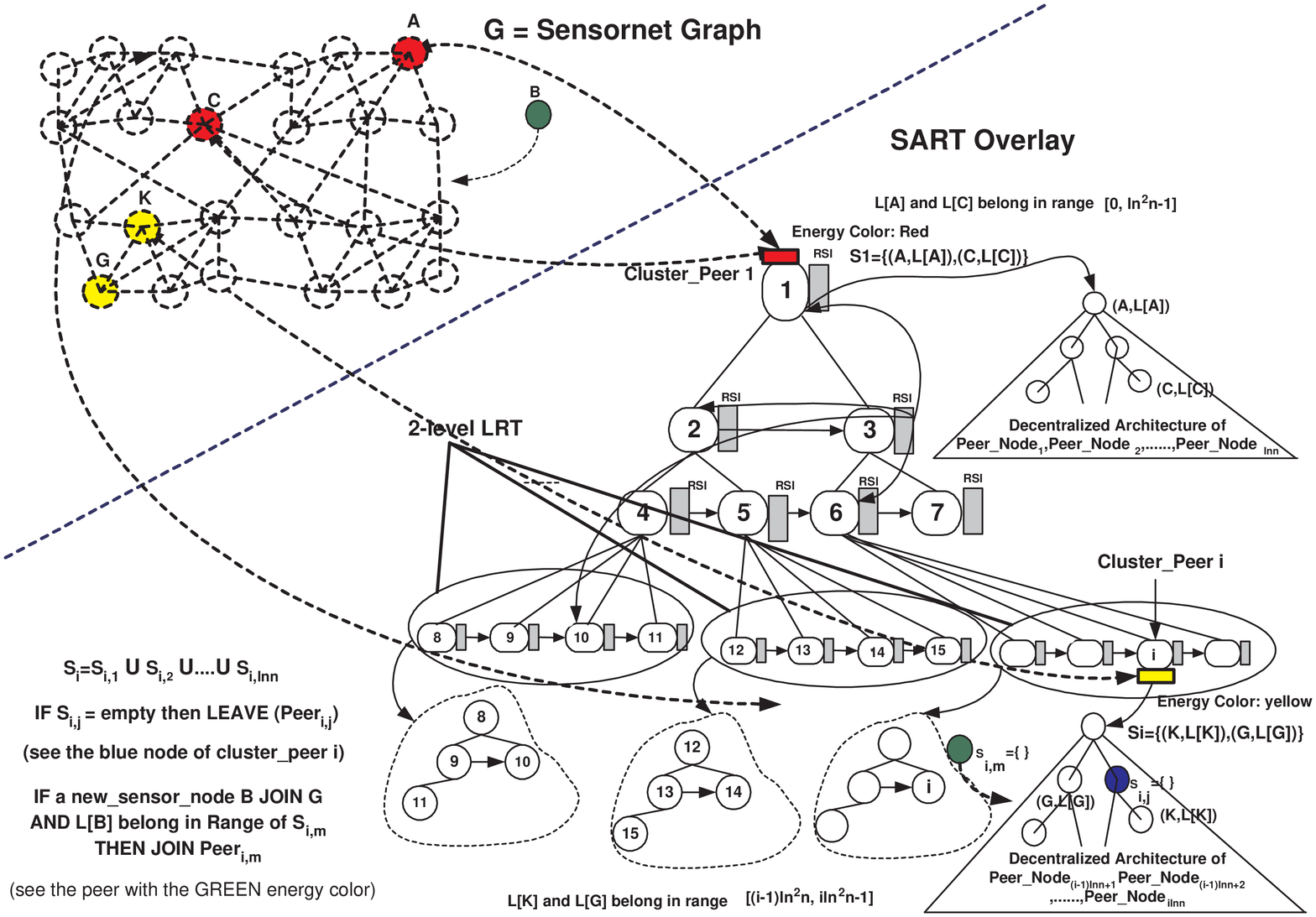}
\end{center}
\caption{Building the SART Bipartite P2P Overlay}
\label{fig:ART}
\end{figure*}

\emph{\textbf{Lookup Algorithms:}} Since the maximum number of nesting levels is $O(\log_{b} \log N)$ and at each nesting level $i$ the standard LRT structure has to be applied in $N^{1/2^i}$ collections, the whole searching process requires $O(\log_{b}^2 \log N)$ hops. Then, the target peer can be located by searching the respective decentralized structure. Through the poly-logarithmic load of each cluster\_peer, the total query complexity $O(\log_{b}^2 \log N)$ follows. Exploiting now the order of keys on each peer, range queries require $O(\log_{b}^2 \log N+\left| A \right|)$ hops, where $\left| A \right|$ the answer size. 

\emph{\textbf{Join/Leave Operations:}} A peer $u$ can make a join/leave request at a
particular peer $v$, which is located at cluster\_peer $W$. Since the size of $W$ is bounded by a $polylogN$ size
in expected w.h.p. case, the peer join/leave can be carried out in $O(loglogN)$ hops.

\emph{\textbf{Node Failures and Network Restructuring:}} Obviously, node failure and network restructuring operations are according to the decentralized architecture used in each cluster\_peer.


\subsection{Building the SART Overlay}

Let $P_{i,j}$ the $j^{th}$ peer of cluster\_peer i. Each overlay peer $P_{i,j}$, stores a set $S_{i,j}=\left\{(g, L[g])\right\}$, where $L[g]$ is a $k-bit$ key belonging in universe $K=[0,2^{k}-1]$, which represents the current energy-level of the sensor node $g$. In particular (and based on design analysis of previous section) it holds that $L[g] \in \left[ (i-1)ln^2n, iln^2n-1\right]$.  Thus, the total set of $Cluster\_Peer$ $i$ becomes $S_i=S_{i,1}\cup S_{i,2} \cup\ldots \cup S_{i,\Theta(polylogN)}$, where $\left|S_{i,j}\right|\leq n$.

For example in Figure ~\ref{fig:ART}, $S_1=\left\{ (A,L[A]), (C, L[C]) \right\}$ is the set of cluster\_peer 1, which stores the
energy-level keys of red (energy color) sensors $A$ and $C$ as well as $S_i=\left\{ (K,L[K]), (G, L[G]) \right\}$ is the set of cluster\_peer i, which stores the energy-level keys of yellow sensors $K$ and $G$. Tuples $(A,L[A])$ and $(C, L[C])$ are located in different peers of the decentralized structure associated to cluster\_peer 1. The same holds for the tuples $(K,L[K])$ and $(G, L[G])$ in the decentralized structure associated to cluster\_peer i.  

According to the complexity analysis of ART structure, the theorem 1 follows: 

\textbf{Theorem 1}: Assume a $SART$ lookup $P2P$ system for the sensor network $G$. The queries of the form (a), (b) and (c) require $O(\log_{b}^2 \log N)$ expected w.h.p. number of messages. The queries of the form (d) and (e) require $O(\log \log N)$ expected w.h.p. number of messages. \\
Let $G$ the sensor network and $T$ the $SART$ overlay. We are located at sensor node $S \in G$ with low energy level $k'$ and we are looking for a sensor node $R \in G$ with the desired energy level $k$. Algorithm 1 depicts the pseudocode for the Sensor\_Net\_Exact\_Match\_Search routine.\\
Let $G$ the sensor network and $T$ the $SART$ overlay. We are located at sensor node $S \in G$ with low energy level $k'$ and we are looking for a sensor node $R \in G$ the desired energy level of which belongs in the range $[k_1, k_2]$. Algorithm 2 depicts the pseudocode for the Sensor\_Net\_Range\_Search routine.\\
Let $G$ the sensor network and $T$ the overlay structure. We are located at sensor node $S \in G$, the energy level of which has been decreased from $k1$ to $k2$. We have to find the new overlay peer to which the update node $S$ is going to be associated. Algorithm 3 depicts the pseudocode for the update\_overlay\_peer routine.\\
Let $G$ the sensor network and $T$ the overlay structure. If a new sensor node $B$ joins $G$ and $L[B] \in S_{i,m}$ then JOIN $P_{i,m}$ (see the peer with the green energy color). Algorithm 4 depicts the respective pseudocode.\\
Let $G$ the sensor network and $T$ the overlay structure. If $S_{i,j}=\oslash$ then LEAVE $P_{i,j}$ (see the blue node of cluster peer i). Algorithm 5 depicts the respective pseudocode.

\begin{algorithm}
\caption{Sensor\_Net\_Exact\_Match\_Search($G$,$S$,$T$,$k'$,$k$,$R$)}
\label{alg_sensornet_exactmatch_search}
\begin{algorithmic}[1]
\STATE Find the peer node to which sensor S (of enerfy level k') is associated;
\STATE Let $p \in T$ the respective overlay peer; 
\STATE $r=send\_overlay\_search(T,p,k)$; \COMMENT {it is the basic lookup routine of ART structure T}
\STATE Let $r \in T$ the peer node which stores sensor nodes 
     with the desired energy-level k and let say R a randomly chossen one;
\STATE Return R

\end{algorithmic}
\end{algorithm}

\begin{algorithm}
\caption{Sensor\_Net\_Range\_Search($G$,$S$,$T$,$k'$,$k_1$,$k_2$,$R$)}
\label{alg_sensornet_range_search}
\begin{algorithmic}[1]
\STATE Find the peer to which sensor S (of enerfy level k') is associated;
\STATE Let $p \in T$ the respective overlay peer; 
\STATE $r=send\_overlay\_range\_search(T,p,k)$; \COMMENT {it is the range searching routine of ART structure T}
\STATE Let $A$ the set of peers the desired energy-level of which belong in range $[k_1, k_2]$ and let say R a randomly chossen one;
\STATE Return R

\end{algorithmic}
\end{algorithm}

\begin{algorithm}
\caption{Update\_Overlay\_Peer($G$,$T$,$S$,$k_1$,$k_2$)}
\label{Update_Overlay_Peer}
\begin{algorithmic}[1]
\STATE Find the peer to which $S$ is associated according to old energy level $k_1$;
\STATE Let $p \in T$ the respective overlay peer; 
\STATE Delete $(S,k_1)$ from $p$;
\STATE $r=send\_overlay\_search(T,p,k_2)$;  
\STATE Insert the tuple $(S,k_2)$ into $r$;

\end{algorithmic}
\end{algorithm}

\begin{algorithm}
\caption{Join\_Overlay\_Peer($G$,$T$,$B$,$L[B]$)}
\label{Join_Overlay_Peer}
\begin{algorithmic}[1]
\STATE Let $L[B] \in S_{i,m}$ and the $m^{th}$ peer of cluster\_peer i does not exist;
\STATE $send\_join\_peer(T,P_{i,m})$; \COMMENT {it is the Join routine of ART structure T}  
\STATE Let $S_{i,m}=\oslash$ the initial empty set of the new inserted peer $P_{i,m}$;
\STATE Insert the tuple $(B,L[B])$ into $S_{i,m}$;

\end{algorithmic}
\end{algorithm}

\begin{algorithm}
\caption{Leave\_Overlay\_Peer($G$,$T$,$P_{i,j}$)}
\label{Leave_Overlay_Peer}
\begin{algorithmic}[1]
\STATE Let $S_{i,j}=\oslash$ the empty set of peer $P_{i,j}$;
\STATE $send\_Leave\_peer(T,P_{i,j})$; \COMMENT {it is the Leave routine of ART structure T}  

\end{algorithmic}
\end{algorithm}

\section{Experiments}

\begin{figure*}
	\centering
		\includegraphics[width=1.00\textwidth]{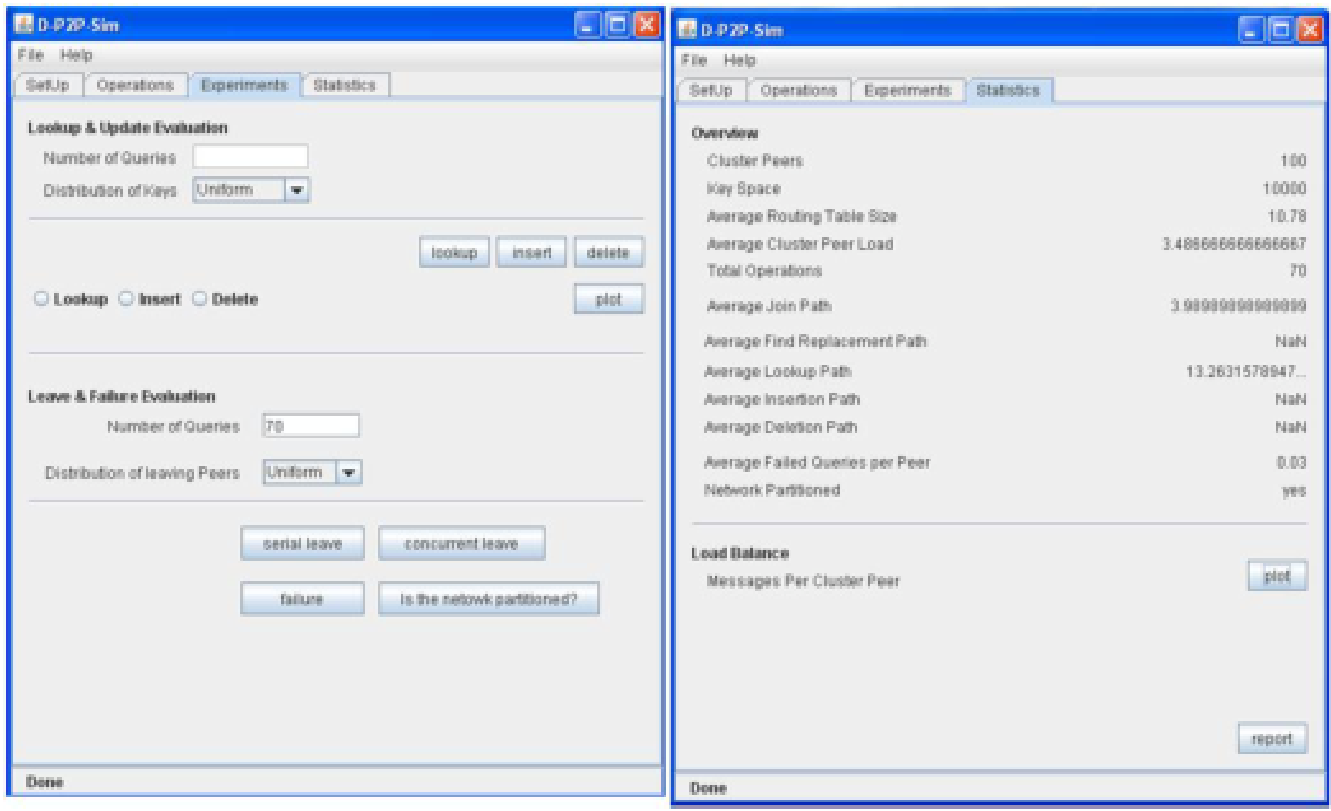}
	\caption{D-P2P-Sim GUI}
	\label{fig:D-P2P-Sim-Gui}
\end{figure*}

For evaluation purposes we used the Distributed Java D-P2P-Sim simulator presented in \cite{SPSTMT10}. The D-P2P-Sim simulator is extremely efficient delivering $>100,000$ cluster peers in a single computer system, using 32-bit JVM 1.6 and 1.5 GB RAM and full D-P2P-Sim GUI support. When 64-bit JVM 1.6 and ~5 RAM is utilized the D-P2P-Sim simulator delivers $>500,000$ cluster peers and full D-P2P-Sim GUI support in a single computer system.

\begin{figure*}
	\centering
		\includegraphics[width=0.70\textwidth]{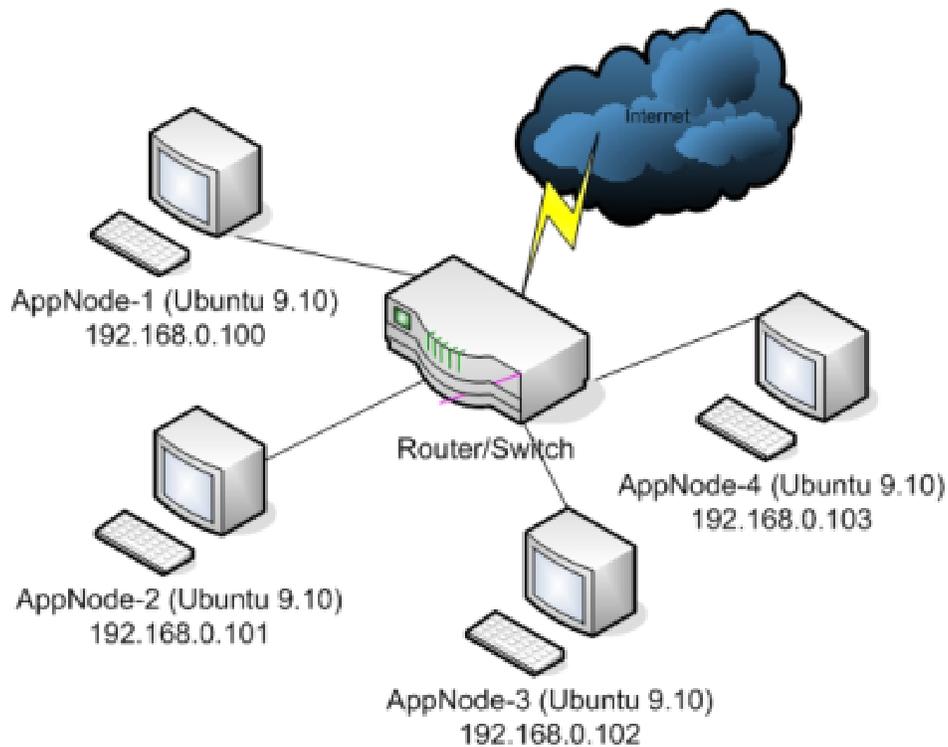}
	\caption{The Distributed Environment}
	\label{fig:distributed-environment}
\end{figure*}

\begin{figure}[htbp]
\centering
		\includegraphics[width=1.00\textwidth]{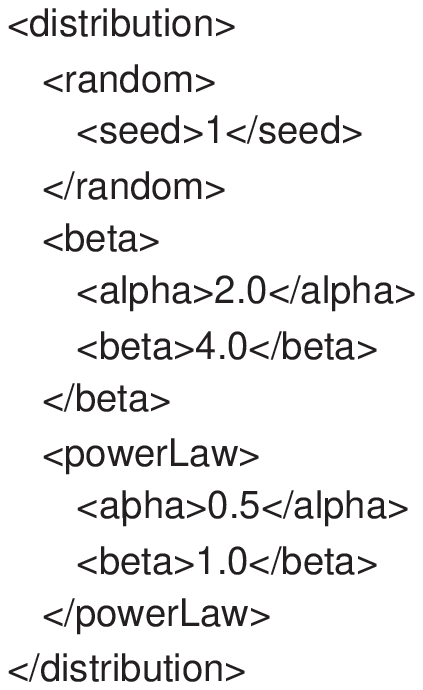}
	\caption{Snippet from config.xml with the pre-defined distribution's parameters setup}
	\label{fig:SART-Configuration}
\end{figure}

The Admin tools of D-P2P-Sim GUI (see Figure \ref{fig:D-P2P-Sim-Gui}) have specifically been designed to support \textsl{reports} on a collection of wide variety of metrics including, protocol operation metrics, network balancing metrics, and even server metrics. Such metrics include frequency, maximum, minimum and average of: number of hops for all basic operations (lookup-insertion-deletion path length), number of messages per node peer (hotpoint-bottleneck detection), routing table length (routing size per node-peer) and additionally detection of network isolation (graph separation). All metrics can tested using a number of different distributions (e.g. normal, weibull, beta, uniform etc). Additionally, at a system level memory can also be managed in order to execute at low or larger volumes and furthermore execution time can also be logged. The framework is open for the protocol designer to introduce additional metrics if needed. Futhermore, XML rule based \textsl{configuration} is supported in order to form a large number of different protocol testing scenarios. It is possible to configure and schedule at once a single or multiple experimental scenarios with different number of protocol networks (number of nodes) at a single PC or multiple PCs and servers distributedly. In particular, when D-P2P-Sim simulator acts in a distributed environment (see Figure \ref{fig:distributed-environment}) with multiple computer systems with network connection delivers multiple times the former population of cluster peers with only ~10\% overhead.


Our experimental performance studies include a detailed performance comparison with TChord, one of the state-of-the-art P2P overlays for sensor networks. Moreover, we implemented each cluster\_peer as a BATON* \cite{JOTVZ06}, the best known decentralized tree-architecture. We tested the network with different numbers of peers ranging up to 500,000. A number of data equal to the network size multiplied by 2000, which are numbers from the universe [1..1,000,000,000] are inserted to the network in batches. The synthetic
data (numbers) from this universe were produced by the following
distributions: beta\footnote{http://acs.lbl.gov/software/colt/api/cern/jet/random/Beta.html}, uniform\footnote{http://docs.oracle.com/javase/1.4.2/docs/api/java/util/Random.html} and power-law\footnote{http://acs.lbl.gov/software/colt/api/cern/jet/random/\\Distributions.html\#nextPowLaw}. The distribution parameters can be easily defined in configuration file\footnote{http://code.google.com/p/d-p2p-sim/downloads/detail?name=Art-config.xml\&can=2\&q=}. Also, the predefined values of these parameters are depicted in the figure
\ref{fig:SART-Configuration}.

For each test, 1,000 exact
match queries and 1,000 range queries are executed, and the average costs
of operations are taken. Searched ranges are created randomly by getting
the whole range of values divided by the total number of peers multiplies
$\alpha$, where $\alpha \in [1..10]$. 
The source code of the whole evaluation process is publicly available \footnote{\tt
http://code.google.com/p/d-p2p-sim/}.

\begin{figure}[htbp]
	\begin{center}
		\includegraphics[scale=0.50]{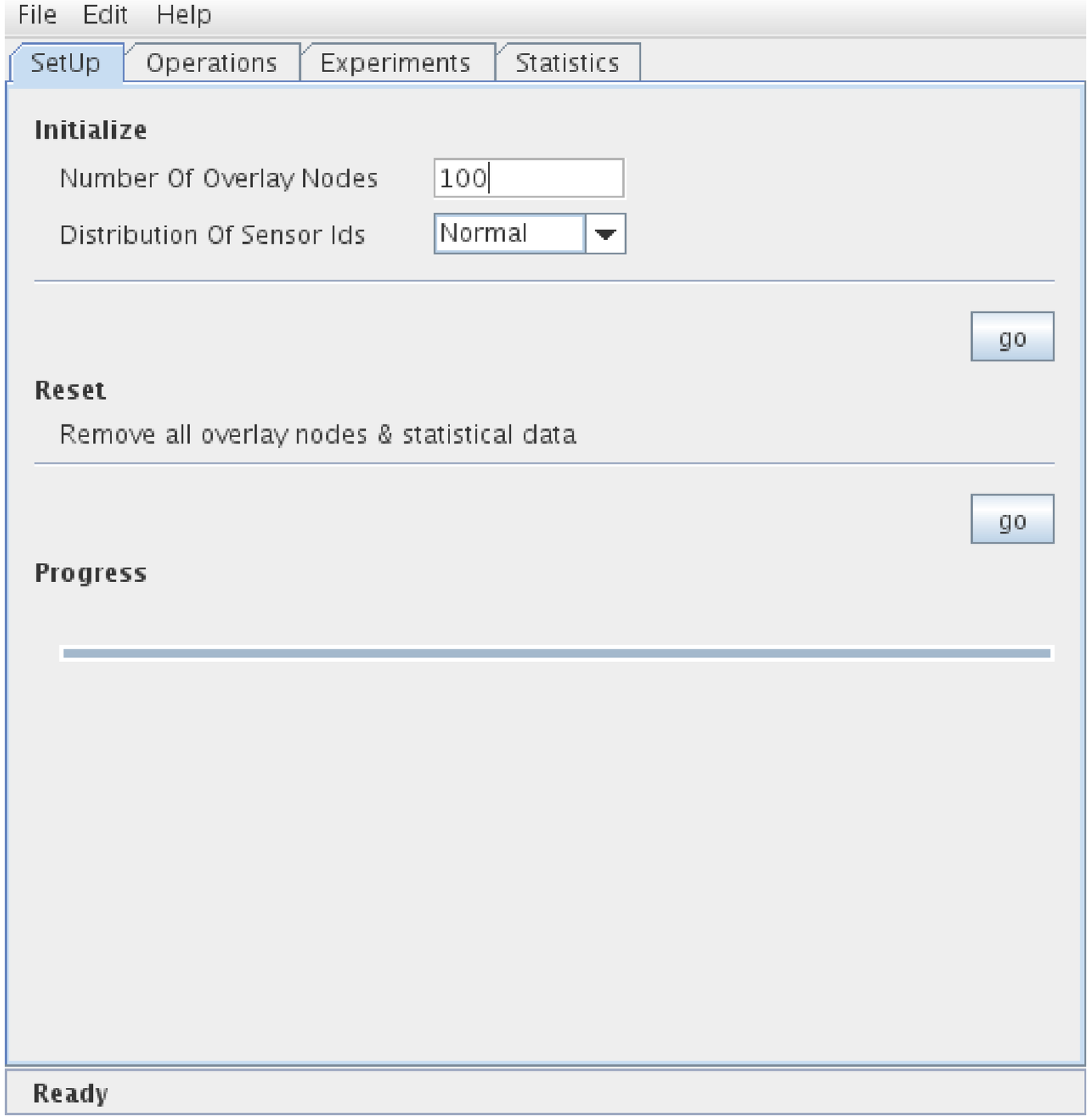}
	\end{center}

	\caption{The tab "SetUp"}
	\label{fig:initialization}
\end{figure}

\begin{figure}[htbp]
	\begin{center}
		\includegraphics[scale=0.50]{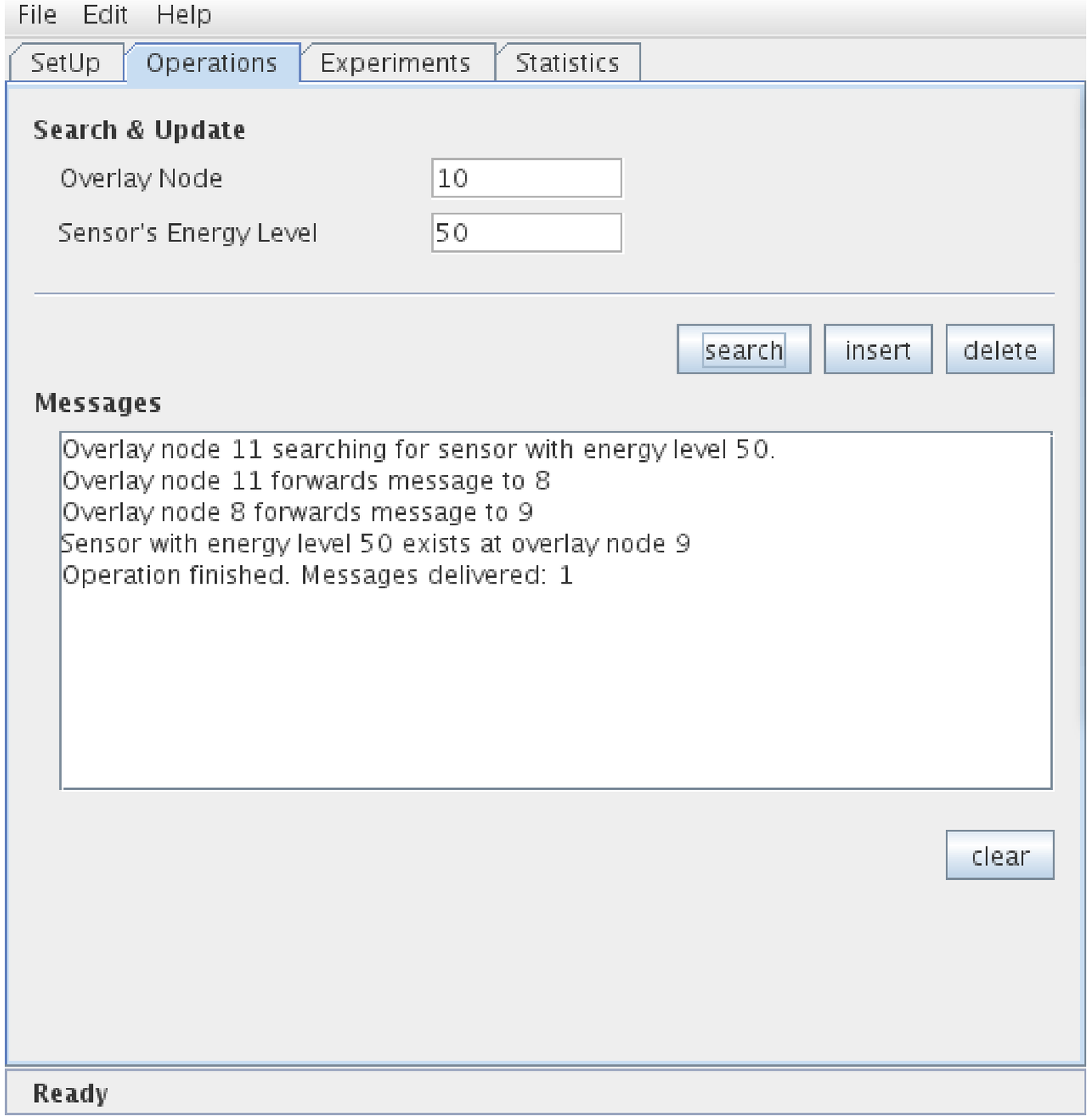}
	\end{center}
	\caption{ The tab "Operations" }
	\label{fig:operations}
\end{figure}

In the first tab (see Figure~\ref{fig:initialization}) the user can set the number of peers which will constitute the overlay and select the energy level distribution over these nodes. The available distributions are: uniform, normal, beta, and pow-law. After the user has set these two fields then the system's initialization can begin.

In the same tab there is a progress bar so the user can obtain the overall process due to the fact that this process may take several minutes. Also there is a button, which resets the system without the need of closing and reopening the simulator if we want to carry out several experiments with different number of peers and energy level distribution.\\

\begin{figure}[htbp]
	\begin{center}
		\includegraphics[scale=0.50]{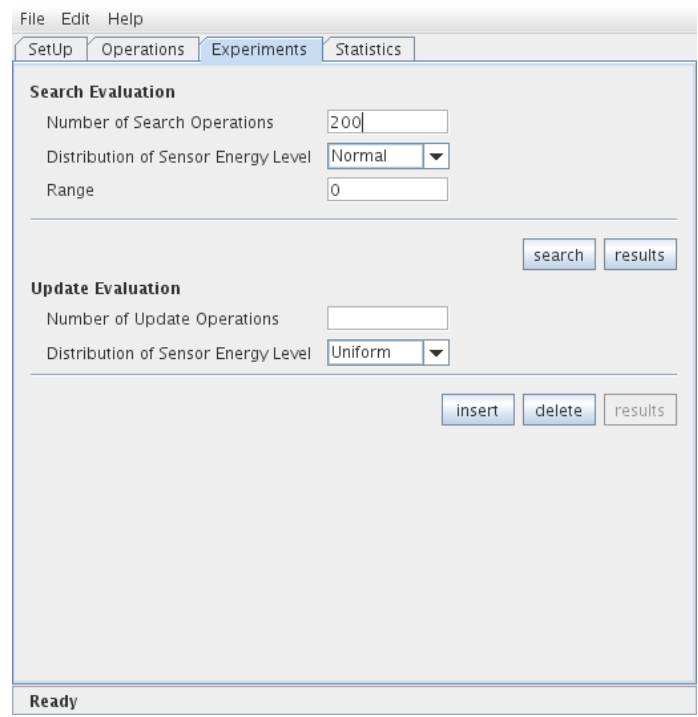}
	\end{center}
	\caption{ The tab "Experiments" }
	\label{fig:experiments}
\end{figure}

The second tab (see Figure~\ref{fig:operations}) provides the ability to search, insert(join) / delete (leave) and update the energy level of a sensor starting the procedure from any peer in the overlay. While one of these operations is being executed, appropriate messages are appearing at the bottom of this tab.






In the third tab (see Figure~\ref{fig:experiments}) the user can prosecute experiments to evaluate the efficiency of the lookup/update operations. There are two panels one for each operation where the user sets the number of the experiments and selects the distribution according to the energy-level keys of the sensors picked up for the experiments. After the termination of the experiments the user can see and save the chart that has been generated. In the forth tab - statistics - the user can see the current number of peers into the system, the number of sensors that have been stored over the peers and the range of sensors' energy level that we can store in the overlay. This tab represents also performance statistics such as the minimum, the maximum and the average path of the total operations that have been performed. Furthermore, this tab generates a chart with the load-balancing over the peers, 
the number of messages that have been forwarded by each peer
)and the number of messages per tree level.


\begin{figure}[htbp]
	\centering
		\includegraphics[width=0.50\textwidth]{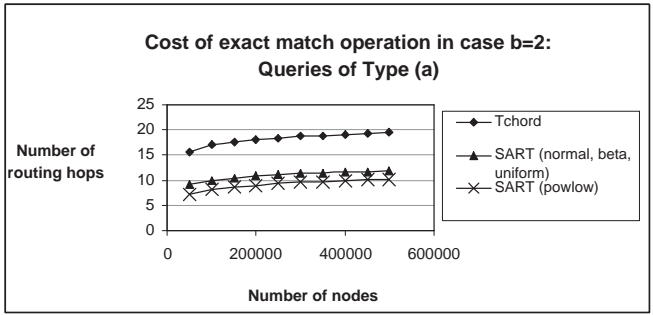}
		\caption{Cost of Exact Match Queries in Case b=2}
	\label{fig:queries-type(a)(b=2)}
\end{figure}

\begin{figure}[htbp]
	\centering
		\includegraphics[width=0.50\textwidth]{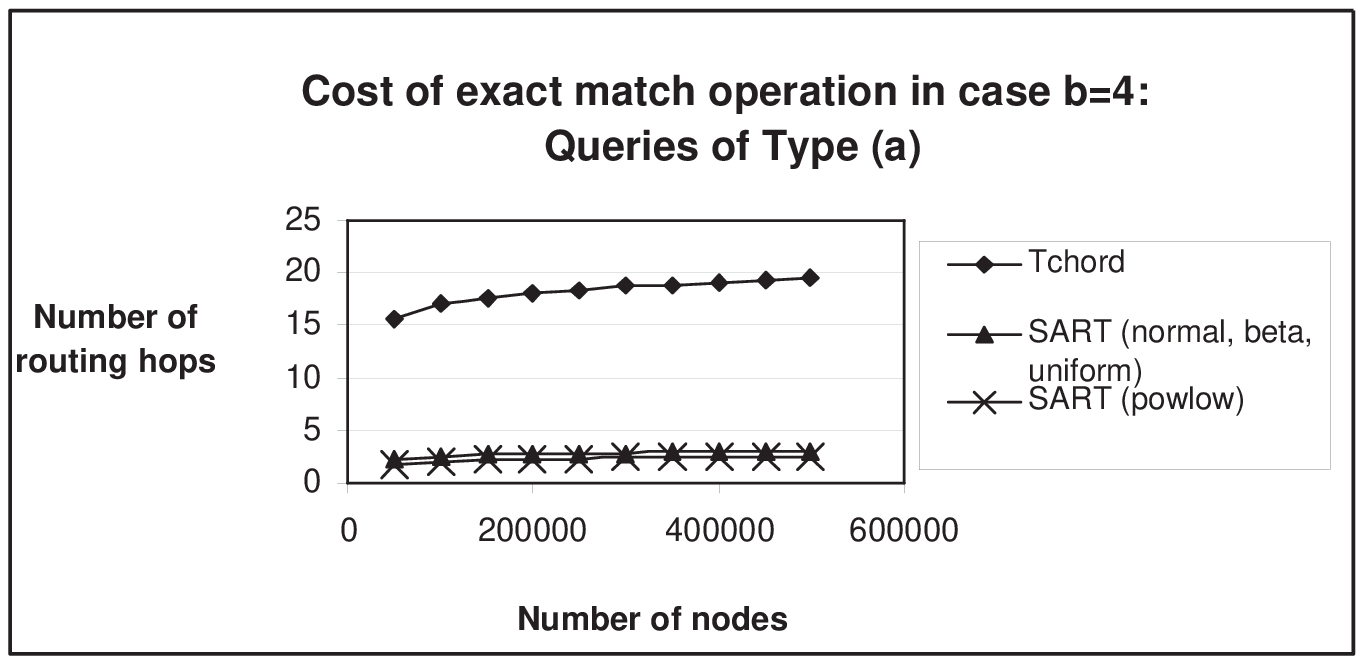}
		\caption{Cost of Exact Match Queries in Case b=4}
	\label{fig:queries-type(a)(b=4)}
\end{figure}

\begin{figure}[htbp]
	\centering
		\includegraphics[width=0.50\textwidth]{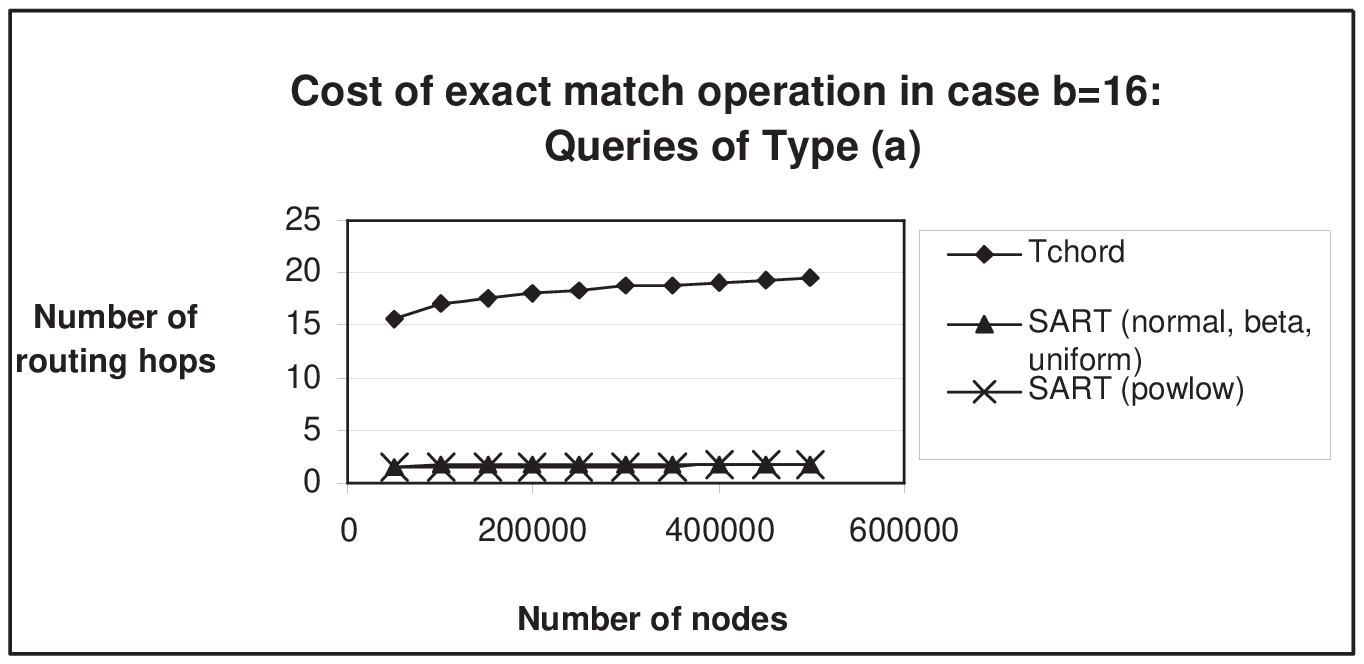}
		\caption{Cost of Exact Match Queries in Case b=16}
	\label{fig:queries-type(a)(b=16)}
\end{figure}


\begin{figure}[htbp]
	\centering
		\includegraphics[width=0.50\textwidth]{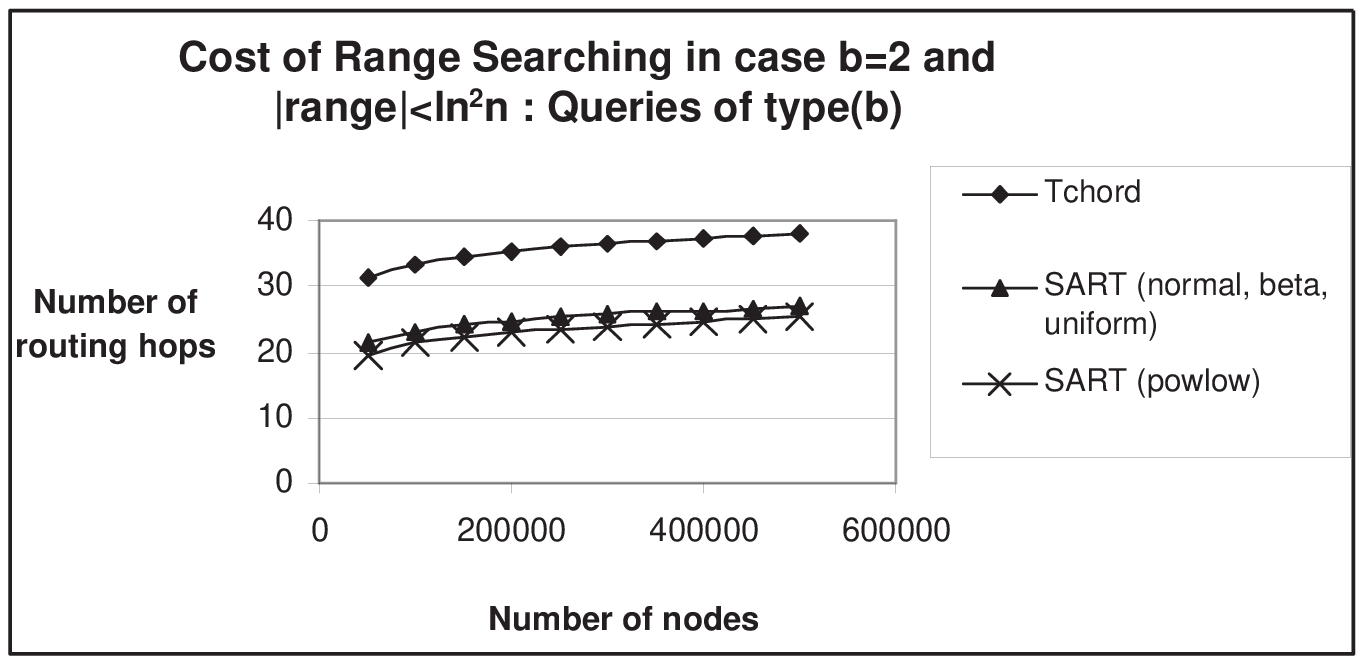}
		\caption{Cost of Range Queries in Case b=2 and $Query\_Range\_Length < Cluster\_Peer\_Key\_Range$}
	\label{fig:queries-type(b)(b=2)(rlecluster)}
\end{figure}

\begin{figure}[htbp]
	\centering
		\includegraphics[width=0.50\textwidth]{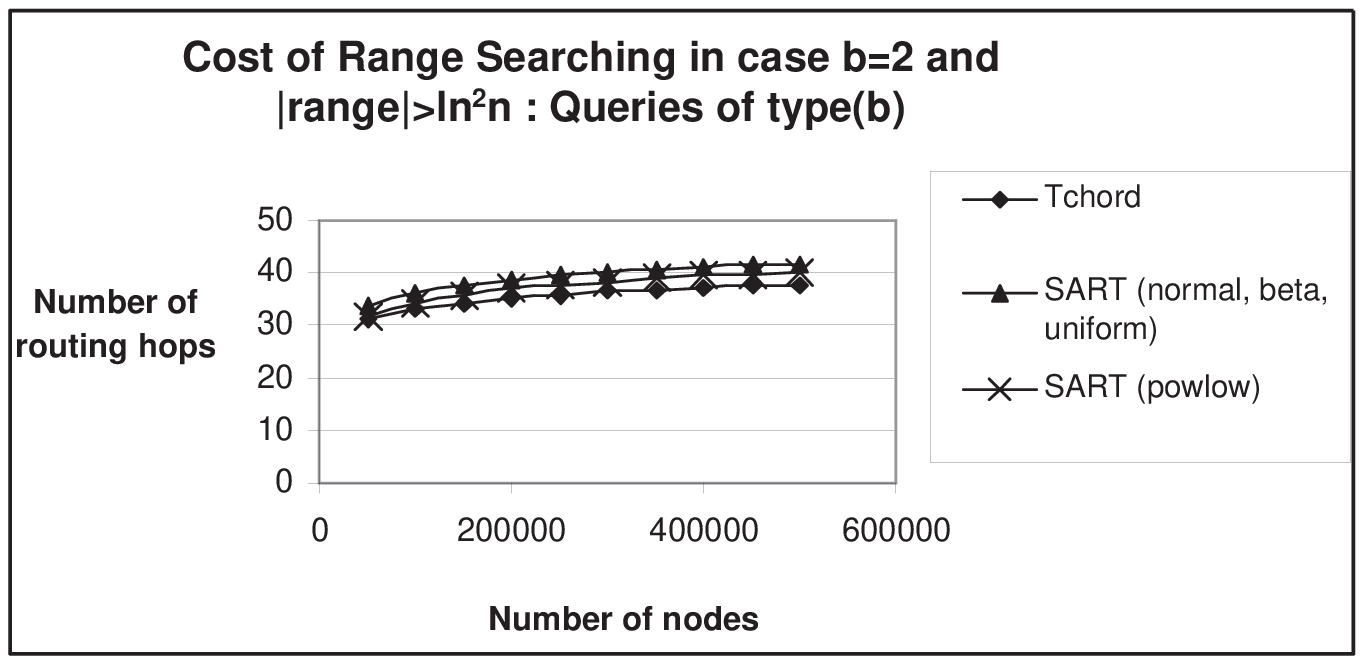}
		\caption{Cost of Range Queries in Case b=2 and $Query\_Range\_Length > Cluster\_Peer\_Key\_Range$}
	\label{fig:queries-type(b)(b=2)(rgecluster)}
\end{figure}
\begin{figure}[htbp]
	\centering
		\includegraphics[width=0.50\textwidth]{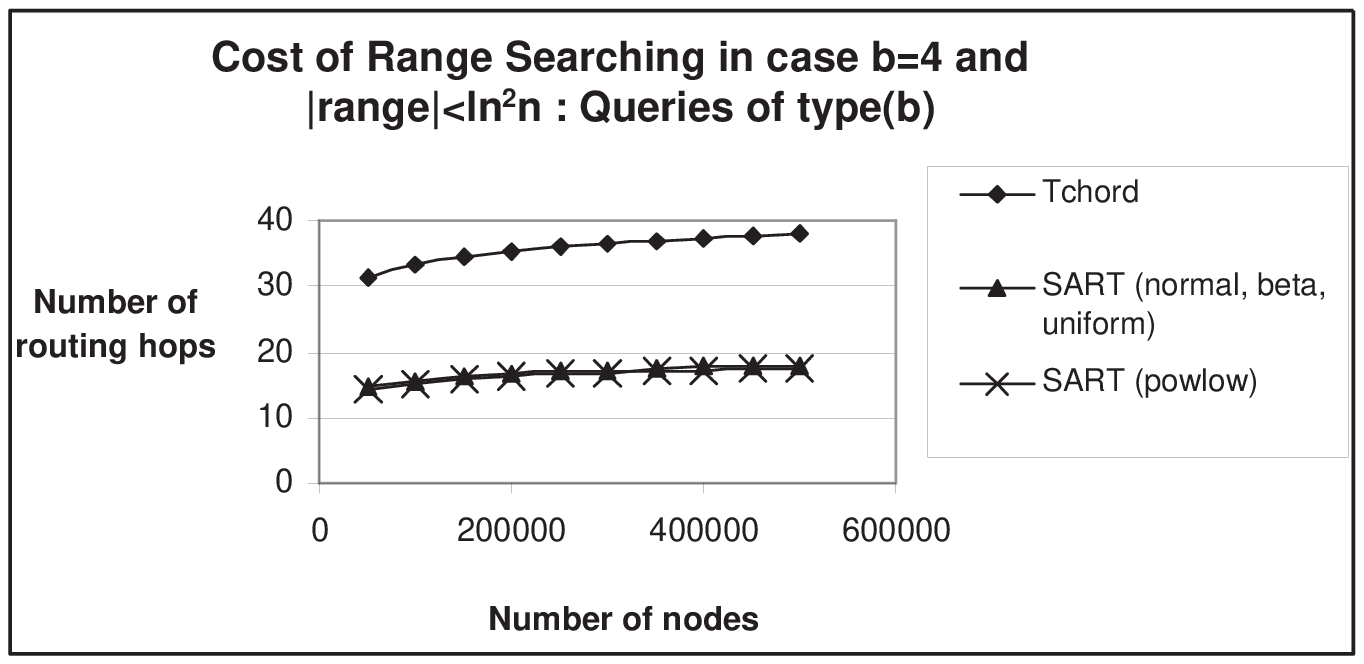}
		\caption{Cost of Range Queries in Case b=4 and $Query\_Range\_Length < Cluster\_Peer\_Key\_Range$}
	\label{fig:queries-type(b)(b=4)(rlecluster)}
\end{figure}

\begin{figure}[htbp]
	\centering
		\includegraphics[width=0.50\textwidth]{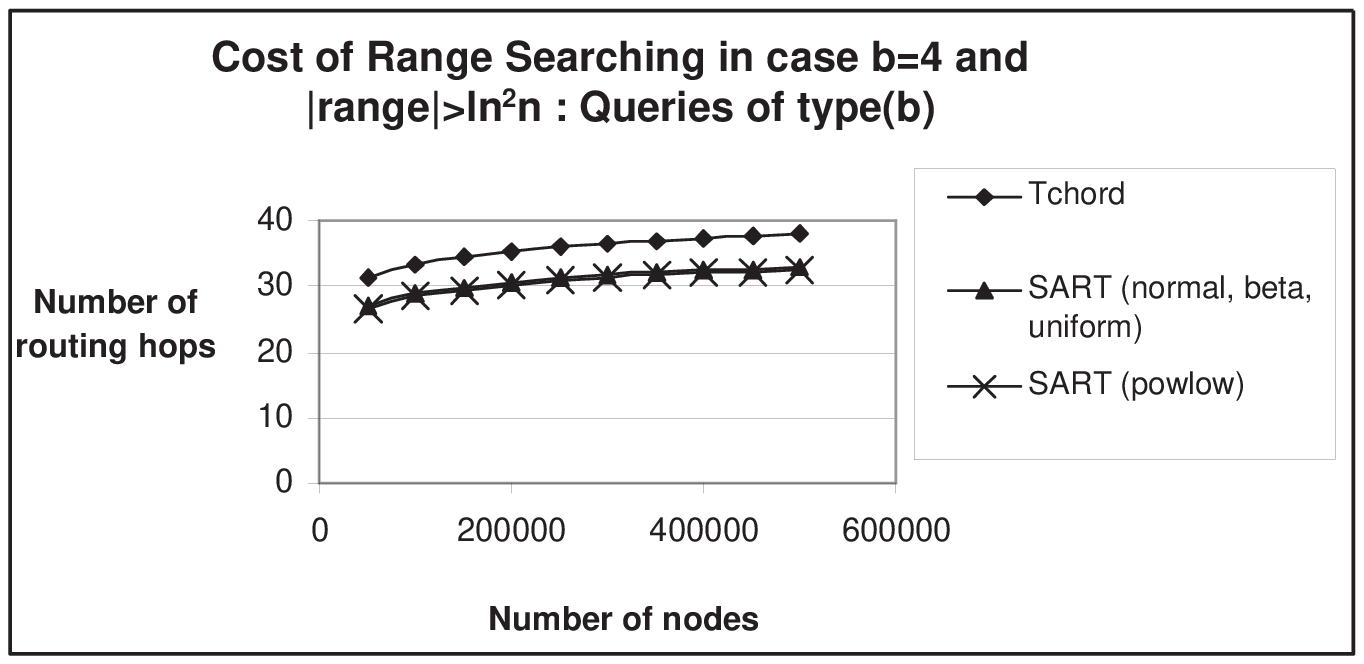}
		\caption{Cost of Range Queries in Case b=4 and $Query\_Range\_Length > Cluster\_Peer\_Key\_Range$}
	\label{fig:queries-type(b)(b=4)(rgecluster)}
\end{figure}

\begin{figure}[htbp]
	\centering
		\includegraphics[width=0.50\textwidth]{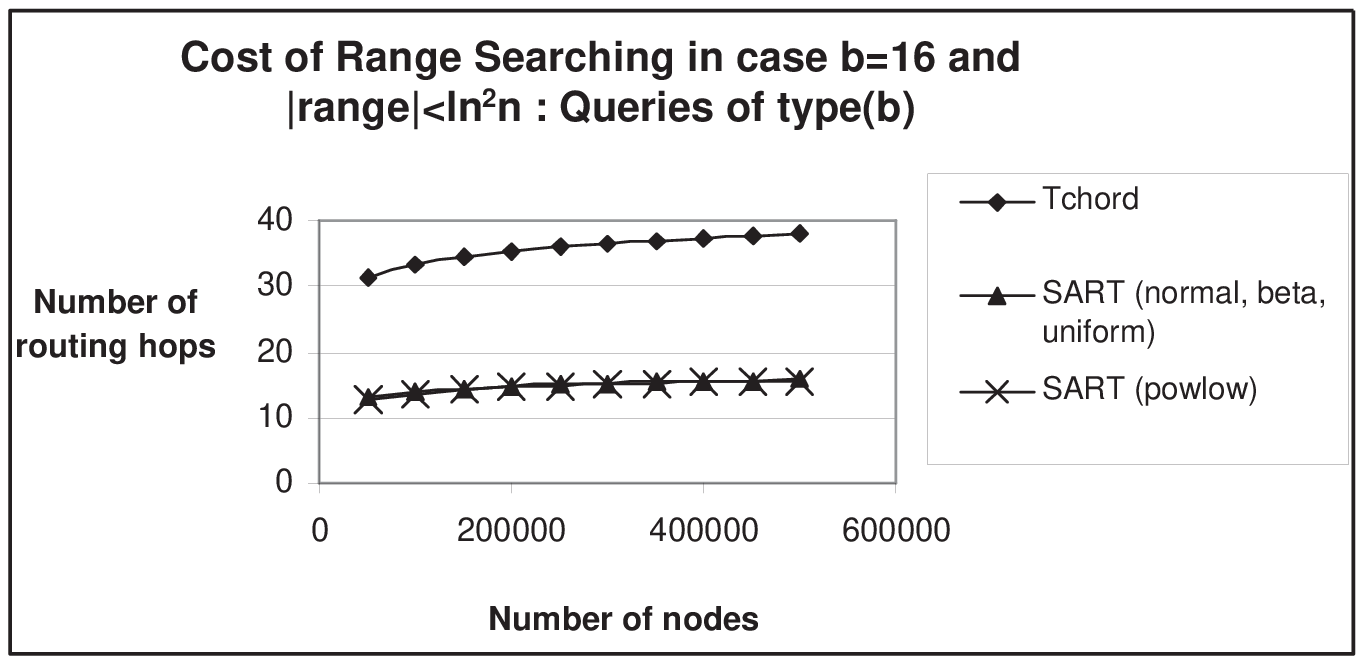}
		\caption{Cost of Range Queries in Case b=16 and $Query\_Range\_Length < Cluster\_Peer\_Key\_Range$}
	\label{fig:queries-type(b)(b=16)(rlecluster)}
\end{figure}

\begin{figure}[htbp]
	\centering
		\includegraphics[width=0.50\textwidth]{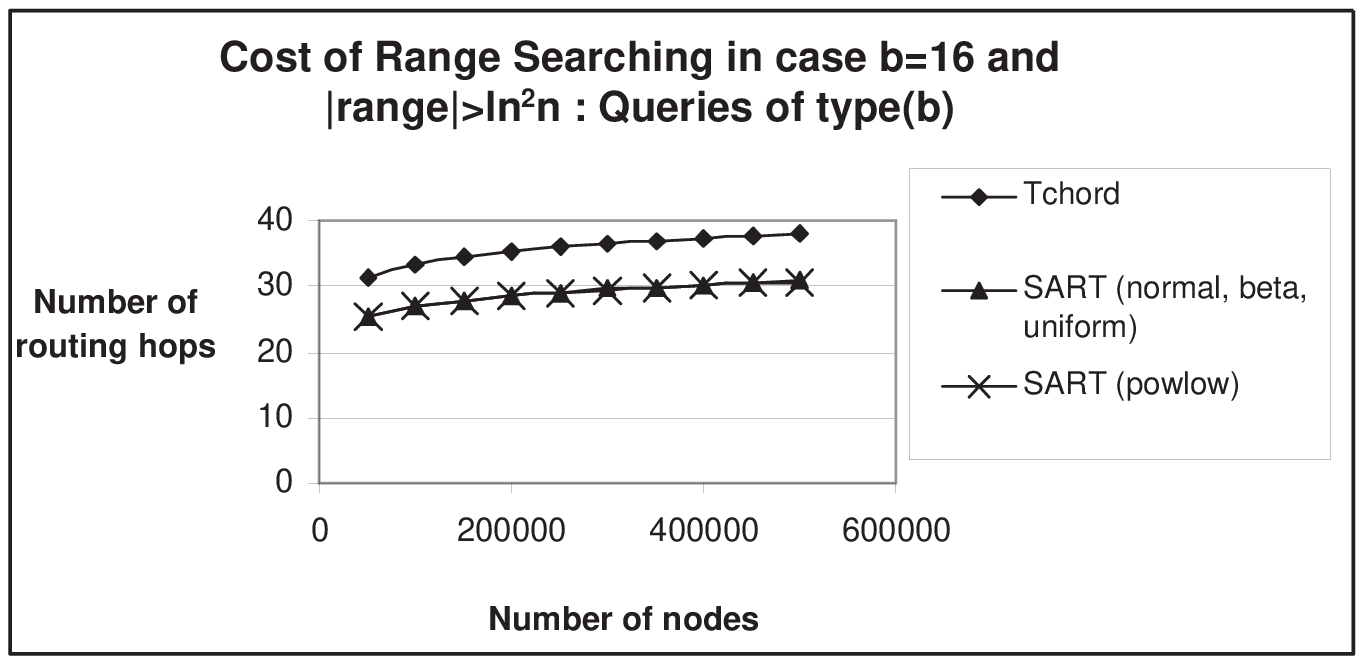}
		\caption{Cost of Range Queries in Case b=16 and $Query\_Range\_Length > Cluster\_Peer\_Key\_Range$}
	\label{fig:queries-type(b)(b=16)(rgecluster)}
\end{figure}


\begin{figure}[htbp]
	\centering
		\includegraphics[width=0.50\textwidth]{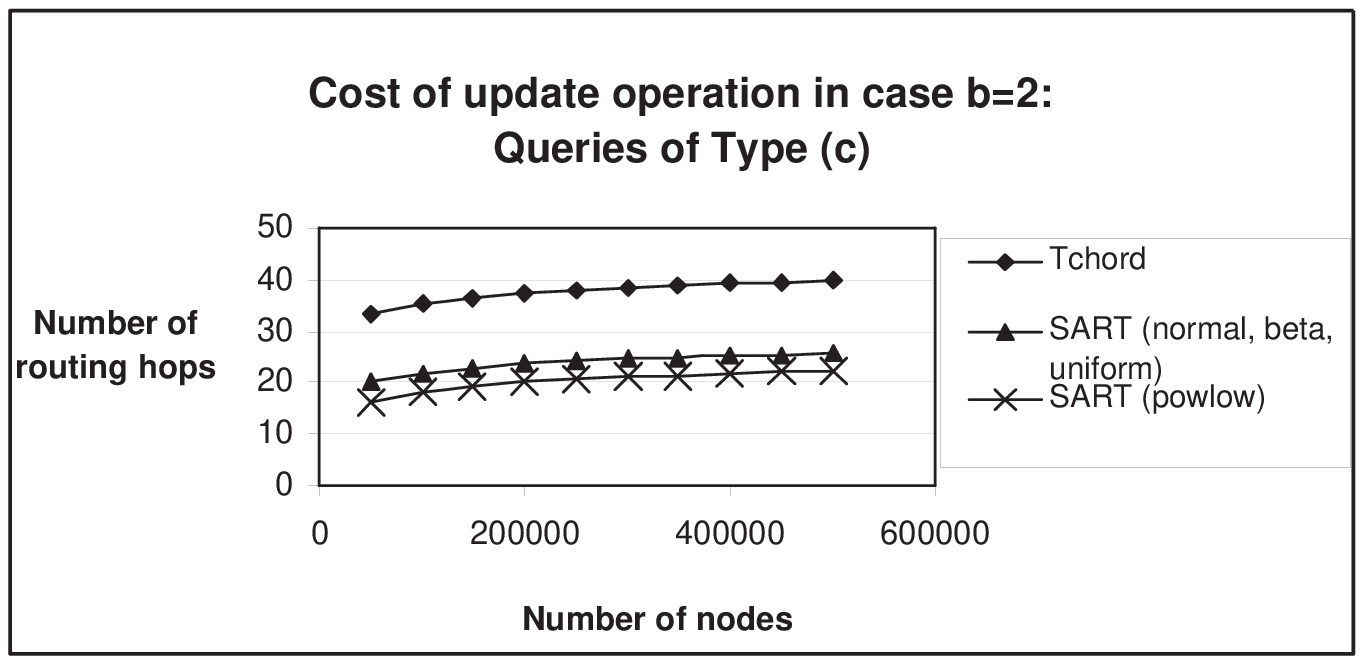}
		\caption{Cost of Update Queries in Case b=2}
	\label{fig:queries-type(c)(b=2)}
\end{figure}

\begin{figure}[htbp]
	\centering
		\includegraphics[width=0.50\textwidth]{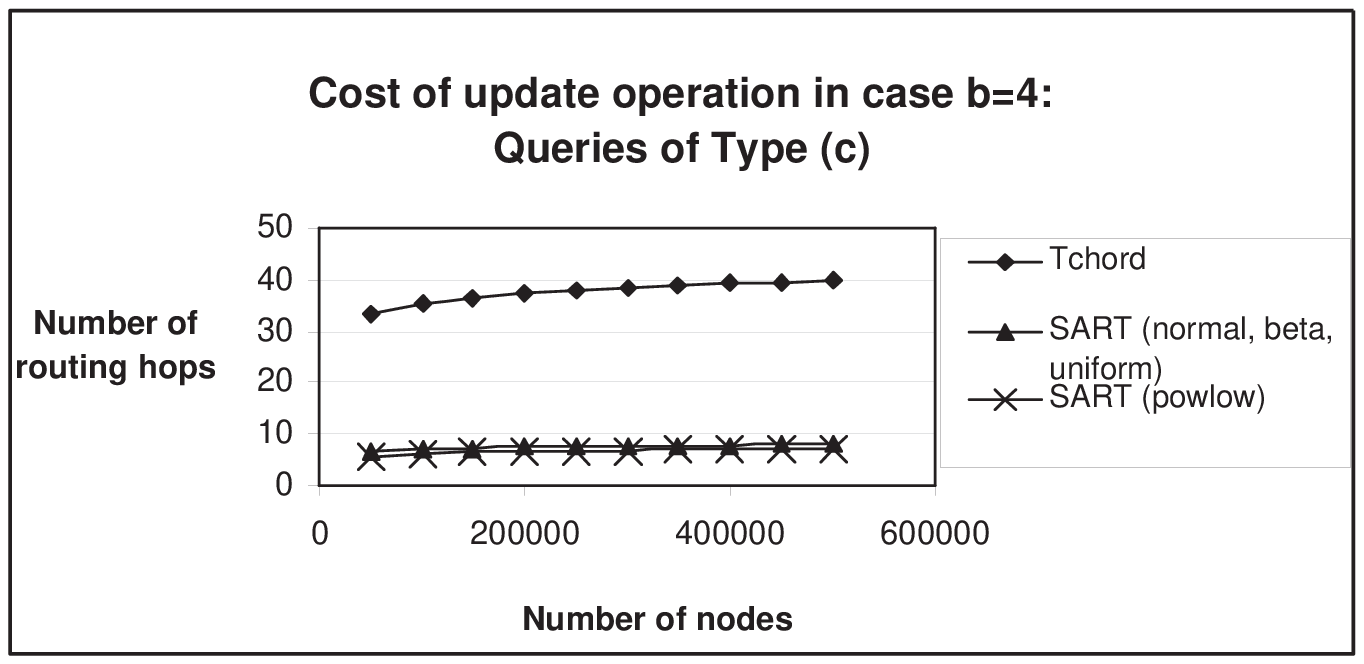}
		\caption{Cost of Update Queries in Case b=4}
	\label{fig:queries-type(c)(b=4)}
\end{figure}

\begin{figure}[htbp]
	\centering
		\includegraphics[width=0.50\textwidth]{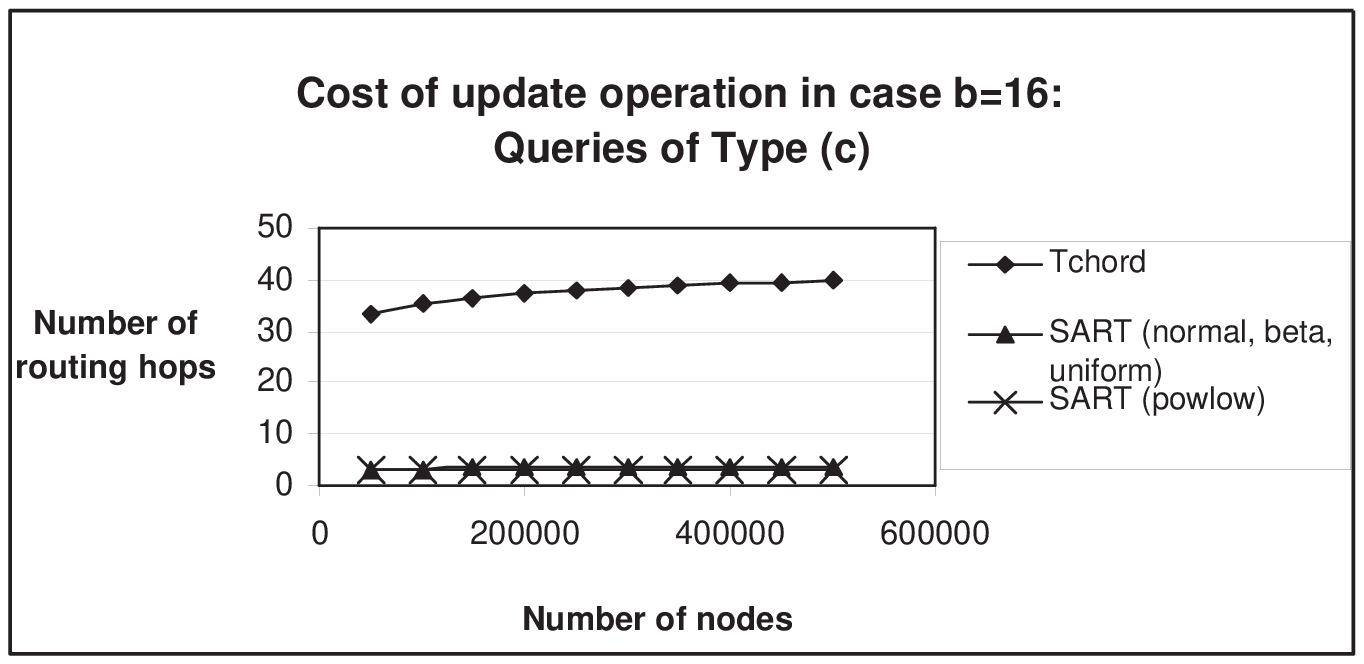}
		\caption{Cost of Update Queries in Case b=16}
	\label{fig:queries-type(c)(b=16)}
\end{figure}

\begin{figure}[htbp]
	\centering
		\includegraphics[width=0.50\textwidth]{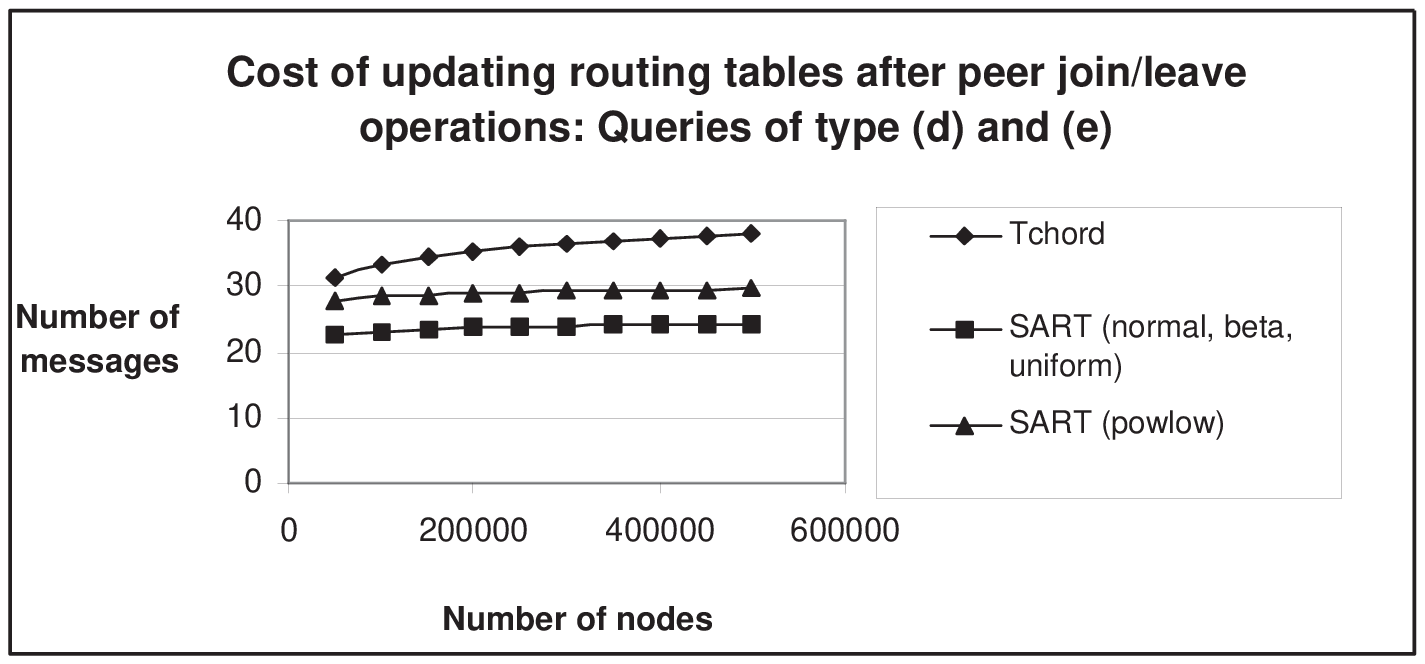}
		\caption{Cost of updating routing tables, after peer join/leave operations: The Cost is independed on parameter $b$}
	\label{fig:queries-type(d,e)}
\end{figure}

In the most of cases, SART outperforms TChord by a wide margin. As depicted in Figures ~\ref{fig:queries-type(a)(b=2)}, \ref{fig:queries-type(a)(b=4)} and \ref{fig:queries-type(a)(b=16)} our method is almost 2 times faster for $b=2$, 4 times faster for $b=4$ and 5 times faster for $b=16$. As a consequence we have a performance improvement from 50\% to 80\%.\\
 
The results are analogous with respect to the cost of range queries as depicted in Figures \ref{fig:queries-type(b)(b=2)(rlecluster)}, \ref{fig:queries-type(b)(b=2)(rgecluster)}, \ref{fig:queries-type(b)(b=4)(rlecluster)}, \ref{fig:queries-type(b)(b=4)(rgecluster)}, \ref{fig:queries-type(b)(b=16)(rlecluster)} and \ref{fig:queries-type(b)(b=16)(rgecluster)}.

In case $Query\_Range\_Length < Cluster\_Peer\_Key\_Range$ and $b=2$, we have an 25\% improvement, however, when $Query\_Range\_Length > Cluster\_Peer\_Key\_Range$, SART and TChord have almost similar performance behaviour.

In case $Query\_Range\_Length < Cluster\_Peer\_Key\_Range$ and $b=4$, we have an 50\% improvement, however, when
$Query\_Range\_Length > Cluster\_Peer\_Key\_Range$ the improvement of our method downgrades to 13.15\%.

In case $Query\_Range\_Length < Cluster\_Peer\_Key\_Range$ and $b=16$, we have an 52.7\% improvement, however, when
$Query\_Range\_Length > Cluster\_Peer\_Key\_Range$ the improvement of our method downgrades to 21.05\%. \\

Figures \ref{fig:queries-type(c)(b=2)}, \ref{fig:queries-type(c)(b=4)} and \ref{fig:queries-type(c)(b=16)} depict the cost of update queries. In particular, for $b=2,4,16$, we have an improvement of 37.5\%, 75\% and 87.5\% respectively. 

Finally, Figure \ref{fig:queries-type(d,e)} depicts the cost of updating the routing tables, after peer join/leave operations. For \emph{bad} or non-smooth distributions, like $powlow$, we have an 23.07\% improvement. However, for more smooth distributions like $beta$, $normal$ or $uniform$ the improvement of our method increases to 38.46\%.

\section{Conclusions}

We considered the problem of constructing efficient P2P overlays for sensornets providing "Energy-Level Application and Services". On this purpose we designed SART, the best-known dynamic P2P overlay providing support for processing queries in a sensornet. We experimentally verified this performance via the D-P2P-Sim framework.

\end{document}